\documentclass[a4paper,10pt,twoside]{cpc-hepnp}

\usepackage{multicol}
\usepackage{graphicx}
\usepackage{booktabs}
\usepackage{longtable}
\usepackage{amssymb,bm,mathrsfs,bbm,amscd}
\usepackage[tbtags]{amsmath}
\usepackage{lastpage}
\usepackage{CJK}

\begin{document}
\begin{CJK*}{GBK}{song}

\fancyhead[c]{\small Submitted to Chinese Physics C} \fancyfoot[C]{\small 010201-\thepage}

\footnotetext[0]{Received 14 March 2009}

\title{Competition between $\alpha$-decay and $\beta$-decay for Heavy and Superheavy Nuclei\thanks{Supported by National Natural Science Foundation of China under Grant No. 11247001, by the Research Foundation of Education
 Bureau of Anhui Province, China under Grant Nos. KJ2012A083 and KJ2013Z066, and by Fundamental Research Funds for
  the Central Universities of China under Grant No. 2012HG2Y0004.}}

\author{%
      SHENG Zong-Qiang$^{1;1)}$\email{zqsheng@aust.edu.cn}%
\quad SHU Liang-Ping$^{1}$
\quad MENG Ying$^{1}$\\
\quad HU Ji-Gang$^{2}$
\quad QIAN Jian-Fa$^{1}$
}
\maketitle

\address{%
$^1$ School of Science, Anhui University of Science and Technology, Huainan 232001, China\\
$^2$ School of Electronic Science and Applied Physics, Hefei University of Technology, Hefei 230009, China\\
}

\begin{abstract}
In this work, the $\beta$-stable region for Z $\geq$
  90 is proposed. The calculated $\beta$-stable nuclei in the
  $\beta$-stable region are in good agreement with the ones
  obtained by M\"{o}ller \emph{et al}.. The half-lives of the nuclei close to the
  $\beta$-stable region are calculated and the competition between
  $\alpha$-decay and $\beta$-decay is systematically investigated. The
  calculated half-lives and the suggested decay modes are well in line
  with the experimental results. The predictions for half-lives and
  decay modes of the nuclei with Z = 107$-$110 are presented.
\end{abstract}

\begin{keyword}
decay, $\beta$-stable region, half-life, superheavy nuclei
\end{keyword}

\begin{pacs}
21.10.Tg, 23.60.+e, 27.90.+b
\end{pacs}

\footnotetext[0]{\hspace*{-3mm}\raisebox{0.3ex}{$\scriptstyle\copyright$}2013
Chinese Physical Society and the Institute of High Energy Physics
of the Chinese Academy of Sciences and the Institute
of Modern Physics of the Chinese Academy of Sciences and IOP Publishing Ltd}%

\begin{multicols}{2}

\section{Introduction}

After Becquerel discovered nuclear radioactivity in
1896, scientists began to investigate nuclear decay and found several
decay modes. $\alpha$-decay and $\beta$-decay are two of the most
important decay modes. $\alpha$-decay was first explained by
Rutherford in 1908~\cite{Ruth}. In the 1930s, Fermi proposed
the basic theory of $\beta$-decay~\cite{Ferm}. From then on,
$\alpha$-decay and $\beta$-decay have been widely
studied theoretically and experimentally~\cite{Hofm,Wilk,Haxt,Koon,Zha3,Ren,Xu,Renyj}, and a large
number of research results and publications have come out. Generally speaking,
$\alpha$-decay mostly occurs in heavy and superheavy nuclei, while
$\beta$-decay can occur throughout the whole periodic Table.

In the early stage of the development of nuclear physics, scientists
could only study the properties of the nuclei very close to the
$\beta$-stable line. As a result, many nuclear phenomena, laws,
formulae, methods, and models were based on the long-lived nuclei or
stable nuclei close to the $\beta$-stable line. It is much easier to
find and synthesize new nuclei close to the $\beta$-stable line.
Nowadays, with the development of radioactive nuclear beams, many
nuclei far from the $\beta$-stable line have been studied. Many new
experimental phenomena have been discovered. At present, the
stable nuclei with $Z<83$ are very clear and definite. On the other hand, most nuclei beyond Z = 83 are unstable, and
their half-lives are usually short. In this article,
the $\beta$-stable region for $Z \geq 90$ will be proposed. The half-lives of the nuclei close to the
$\beta$-stable region will be calculated and the competition
between $\alpha$-decay and $\beta$-decay will be investigated. Then the decay modes can be suggested by the
results of competition.

This article is organized in the following way. In Sec. 2, the
$\beta$-stable region for $Z \geq 90$ is proposed. In Sec. 3, the
half-lives of the nuclei close to the $\beta$-stable
region are calculated and the competition between $\alpha$-decay and
$\beta$-decay is studied. A summary is given in Sec. 4.

\section{The $\beta$-stable region for Z $\geq $ 90}

The $\beta$-stable line for Z $<$ 83 has been well studied
by physicists. For heavy and superheavy nuclei with Z $\geq$ 90, most
of them can occur $\alpha$-decay and $\beta$-decay simultaneously,
and their half-lives are usually short.  For this reason, it is more important to study the $\beta$-stable region than to
study the $\beta$-stable line for these heavy and superheavy nuclei.
In this section, we investigate the boundary of the $\beta$-stable
region based on a successful binding energy formula.

To accurately measure and calculate the ground-state nuclear binding
energies (or masses) is an important goal of nuclear physicists. The
binding energy plays a crucial role for the nuclear stability
on $\beta$-decay, $\alpha$-decay and spontaneous fission of
heavy-mass region with Z $\geq$ 90. In Ref. \cite{Don2}, Dong and
Ren proposed a binding energy formula for heavy and superheavy
nuclei. One can accurately reproduce the binding
energies for the known heavy and superheavy nuclei with this formula. This formula is
useful for accurately estimating the binding energies of unknown
superheavy nuclei. Its form is the following:
\begin{equation}
\begin{array}{lll}
B(Z,A)=&a_{v}A-a_{s}A^{2/3}-a_{c}Z^{2}A^{-1/3} -a_{a}(\dfrac{A}{2}-Z)^{2}A^{-1}\\
&+a_{p}A^{-1/2}+\dfrac{a_{6}|A-252|}{A}-\dfrac{a_{7}|N-152|}{N}\\
&+\dfrac{a_{8}|N-Z-50|}{A}.\\
\end{array}
\end{equation}
The best fit parameters are
\begin{equation}
\left\{
\begin{array}{lll}
a_{v}=&15.8032&MeV,\\
a_{s}=&17.8147&MeV,\\
a_{c}=&0.71478&MeV,\\
a_{a}=&97.6619&MeV,\\
a_{6}=&5.33&MeV,\\
a_{7}=&21.0&MeV,\\
a_{8}=&-15.25&MeV.
\end{array}
\right.
\end{equation}
The coefficients of the pairing energy are
\begin{equation}
a_{p}=
 \left\{
\begin{array}{llll}
12.26&MeV,&even-even\;\,nuclei,\\
3.0&MeV,&even-odd\;\,nuclei,\\
0&MeV,&odd-even\;\,nuclei,\\
-8.0&MeV,&odd-odd\;\,nuclei.
\end{array}
\right.
\end{equation}

The mass formula has the form:
\begin{equation}
\begin{array}{lll}
M(Z,A)&=&ZM_{H}+NM_{n}-B(Z,A)\\
&=&AM_{n}+Z(M_{H}-M_{n})-B(Z,A),
\end{array}
\end{equation}
where $(M_{H} -M_{n}) = -0.782$ MeV.

The decay energies of $\beta^{-}$-decay and $\beta^{+}$-decay can be
written as:
\begin{equation}
E_{d}(\beta^{-})=M(Z,A)-M(Z+1,A),
\end{equation}
\begin{equation}
E_{d}(\beta^{+})=M(Z,A)-M(Z-1,A)-2m_{e},
\end{equation}
where $2m_{e} = 1.022$ MeV.

From the Eqs. (1), (4), (5) and (6), we get:
\begin{equation}
\begin{array}{lll}
E_{d}(\beta^{-})=&0.782-a_{c}(2Z+1)A^{-1/3}+\dfrac{a_{a}(A-2Z-1)}{A}\\
&+a_{7}(\dfrac{|A-Z-152|}{A-Z}-\dfrac{|A-Z-153|}{A-Z-1})\\
&+\dfrac{a_{8}}{A}(|A-2Z-52|-|A-2Z-50|),\\
\end{array}
\end{equation}
\begin{equation}
\begin{array}{lll}
E_{d}(\beta^{+})=&-1.804-a_{c}(-2Z+1)A^{-1/3}-\dfrac{a_{a}(A-2Z+1)}{A}\\
&+a_{7}(\dfrac{|A-Z-152|}{A-Z}-\dfrac{|A-Z-151|}{A-Z+1})\\
&+\dfrac{a_{8}}{A}(|A-2Z-48|-|A-2Z-50|).\\
\end{array}
\end{equation}
If the values of $E_{d}(\beta^{-})$ and $E_{d}(\beta^{+})$ are set
to zero, one can get the limits of $\beta^{-}$-decay and
$\beta^{+}$-decay for each isotopic chain. For each fixed proton
number Z, one can get two different mass numbers for the limits of
$\beta^{-}$-decay and $\beta^{+}$-decay, respectively. For all the
proton numbers from Z = 90 to Z = 126, two sets of mass numbers for
the limits of $\beta^{-}$-decay and $\beta^{+}$-decay can be
obtained, respectively. Connecting two sets of mass numbers for the
limits of $\beta^{-}$-decay and $\beta^{+}$-decay in the coordinate
space (Z, A), the boundary of the limits of
$\beta^{-}$-decay and $\beta^{+}$-decay can be obtained. The calculated results are
plotted in figure 1.

\begin{center}
\includegraphics[width=8cm]{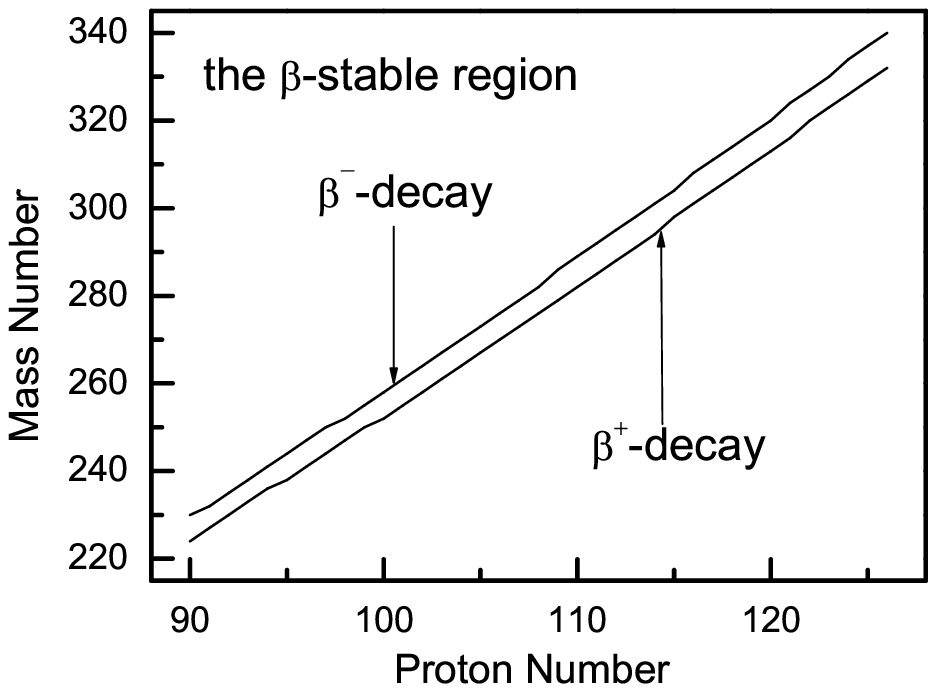}
\figcaption{\label{fig1} The calculated $\beta$-stable region for Z $\geq$ 90. The two curves
denote the limits of $\beta^{-}$-decay and $\beta^{+}$-decay, respectively.}
\end{center}

In figure 1, the two curves denote the limits of
$\beta^{-}$-decay and $\beta^{+}$-decay, respectively. They are
almost parallel. The calculated $\beta$-stable region is a long and
narrow region between the two curves. According to the calculations,
the nuclei above the calculated $\beta$-stable region can occur
$\beta^{-}$-decay and the nuclei below the region can occur
$\beta^{+}$-decay. The nuclei in the $\beta$-stable region are the
possible $\beta$-stable nuclei. Because there are few experimental
data, we compare our calculated results with the results given by
M\"{o}ller \emph{et al}.~\cite{Moll}. The comparison between
our calculated results and M\"{o}ller's results are shown in table
1.
\begin{center}
\tabcaption{ \label{tab1}  The possible $\beta$-stable nuclei in the calculated
$\beta$-stable region for Z $\geq$ 90. The corresponding
$\beta$-stable nuclei calculated by M\"{o}ller \emph{et al}.~\cite{Moll} are listed for comparison.}
\footnotesize
\begin{tabular*}{80mm}{c c l }
\hline
 Z&\multicolumn{2}{c}{Mass number A of $\beta$-stable nuclei}\\
&Cal.&M\"{o}ller\\
\hline
90&224$-$230&224, 226$-$230, 232\\
91&227$-$232&231\\
92&230$-$235&230, 232$-$236, 238\\
93&233$-$238&237\\
94&236$-$241&236, 238$-$242, 244\\
95&238$-$244&241\#, 243\#\\
96&241$-$247&240, 242$-$246, 248\\
97&244$-$250&247\\
98&247$-$252&246, 248$-$252, 254\\
99&250$-$255&253\\
100&252$-$258&252, 254$-$258, 260, 262\\
101&255$-$261&259\\
102&258$-$264&258, 260$-$264, 266\\
103&261$-$267&265\\
104&264$-$270&264, 266$-$268, 270, 272, 274\\
105&267$-$273&269, 271\\
106&270$-$276&268, 270, 272$-$276, 278, 280\\
107&273$-$279&277\\
108&276$-$282&274, 276, 278$-$284, 286\\
109&279$-$286&283\#, 285\#\\
110&282$-$289&282, 284$-$288, 290\\
111&285$-$292&289\\
112&288$-$295&288, 290$-$294, 296\\
113&291$-$298&293\#, 295\#\\
114&294$-$301&292, 294$-$298, 300, 302, 304\\
115&298$-$304&299, 301, 303\\
116&301$-$308&300, 302, 304$-$306, 308, 310$-$312, 314\\
117&304$-$311&307, 309\\
118&307$-$314&304, 306, 308, 310$-$314, 318\\
119&310$-$317&315\\
120&313$-$320&312, 316$-$318, 320, 322, 324\\
121&316$-$324&319\\
122&320$-$327&318, 320, 321, 323$-$326, 328, 330\\
123&324$-$330&322, 327\\
124&327$-$334&317, 323, 324, 326, 328$-$332, 334, 336, 338\\
125&329$-$337&325, 327, 333\\
126&332$-$340&326, 330, 332, 334$-$338\\
\hline
\end{tabular*}
\end{center}

In table 1, the calculated possible $\beta$-stable nuclei and the
results calculated by M\"{o}ller \emph{et al}.~\cite{Moll}are
listed in the second and the third columns, respectively. The mass
numbers with \# denote that the nuclei with these mass numbers are
$\beta$-stable nuclei by estimated from systematic trends in
neighboring nuclei. Our calculated results show that there are
several (from six to nine) $\beta$-stable nuclei in each isotopic
chain and they are continuous in their isotopic chain. For even Z,
the two results are almost the same. On the whole, the range by M\"{o}ller \emph{et al}. is slightly larger
than our calculated results. The calculated $\beta$-stable nuclei
by M\"{o}ller \emph{et al}. are not continuous in their
isotopic chains. Some nuclei are $\beta$-stable nuclei from our
calculations, but they are not $\beta$-stable nuclei from
M\"{o}ller's results, and vice versa. For odd Z, M\"{o}ller's
results show that there are only one or two $\beta$-stable nuclei in
their isotopic chains except for Z = 115, 125. It is different from
our results. But it can be seen that the $\beta$-stable nuclei from
M\"{o}ller's results are all in the middle of our calculated
$\beta$-stable region except for Z = 123, 125. From the above
discussions, it can be said that the calculated $\beta$-stable
region are in good agreement with the M\"{o}ller's results. For
further comprehending the calculated $\beta$-stable region, the
$\beta$-stable nuclei from both the calculated results and M\"{o}ller's results are drawn in figure 2.

\begin{center}
\includegraphics[width=8cm]{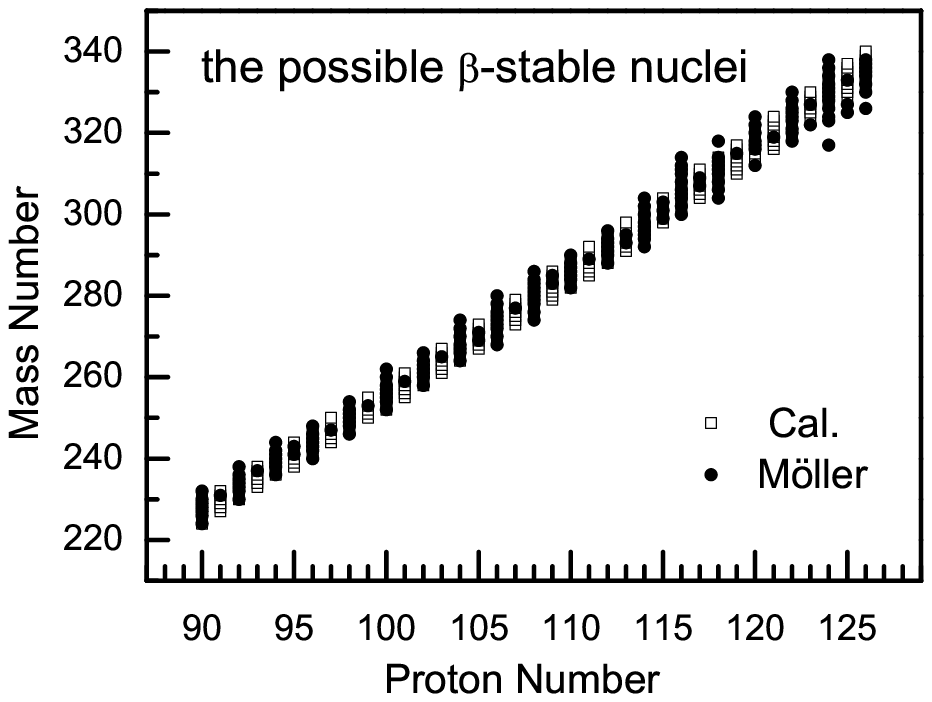}
\figcaption{\label{fig2} The
possible $\beta$-stable nuclei for Z $\geq$ 90. The corresponding
$\beta$-stable nuclei calculated by M\"{o}ller \emph{et al}.~\cite{Moll} are listed for comparison.}
\end{center}

In figure 2, the hollow squares denote the possible $\beta$-stable
nuclei obtained by our calculations, and the dark circles are the
ones from M\"{o}ller's results. For even Z, the $\beta$-stable
nuclei obtained from M\"{o}ller's results almost cover the
calculated $\beta$-stable nuclei. For odd Z, except Z = 123, 125,
the $\beta$-stable nuclei obtained from M\"{o}ller's results are
included in the calculated $\beta$-stable nuclei. It is in line with
the above discussions.

\section{Competition between $\alpha$-decay and $\beta$-decay of the nuclei close to the $\beta$-stable region}

In the previous section, the $\beta$-stable region for Z
$\geq$ 90 has been proposed. Most nuclei with Z $\geq$ 90 can occur
$\alpha$-decay, $\beta$-decay and spontaneous fisson simultaneously. In
this section, we will calculate the half-lives of the nuclei close to the
calculated $\beta$-stable region, and study the competition between
$\alpha$-decay and $\beta$-decay of them. It is a very interesting
topic. There are plenty of experimental half-lives and the decay
modes of many nuclei are very explicit in this region. The
calculated results can be compared with these experimental data and
the reliability of the calculated results can be tested. On the
other hand, the predictions are useful for quickly estimating the decay
modes and half-lives of future superheavy experiments. Before
calculating the half-lives, we firstly introduce several successful
formulae for calculating.

$\alpha$-decay is a very general decay mode for the ground states of
heavy and superheavy nuclei. In Ref.~\cite{Ni}, Ni \emph{et al}.
proposed a unified formula of half-lives for $\alpha$-decay and
cluster radioactivity. For $\alpha$-decay, it is written as:
\begin{equation}
\texttt{log}_{10}T_{1/2}=2a\sqrt{\mu}(Z-2)Q_{\alpha}^{-1/2}+b\sqrt{\mu}
\;[2(Z-2)]^{1/2}+c,
\end{equation}
where $\mu= 4(A-4)/A$, $T_{1/2}$ is the half-life of $\alpha$-decay
(in seconds), and $Q_{\alpha}$ is $\alpha$-decay energy (in MeV). A
and Z are the mass number and the proton number of the parent nuclei
respectively. The values of the parameters are $a=0.39961$,
$b=-1.31008$. Parameter \emph{c }is determined to be $c_{e-e} =
-$17.00698 (for even-even nuclei), $c_{e-o} = -$16.26029 (for
even-odd nuclei), $c_{o-e} = -$16.40484 (for odd-even nuclei), and
$c_{o-o} = -$15.85337 (for odd-odd nuclei).

$\beta$-decay is also a very important decay mode for rich-neutron
or rich-proton nuclei. For $\beta^{-}$-decay, in Ref.~\cite{Zhan},
Zhang \emph{et al}. proposed a reliable formula to calculate the
$\beta^{-}$-decay half-lives. It is written as:
\begin{equation}
\texttt{log}_{10}T_{1/2}=(c_{1}Z+c_{2})N+c_{3}Z+c_{4}+\texttt{shell}(Z,N),
\end{equation}
where
\begin{equation}
\begin{array}{lll}
\texttt{shell}(Z,N)=&c_{5}[e^{-(N-29)^{2}/15}+e^{-(N-50)^{2}/37}\\
&+e^{-(N-85)^{2}/9}+e^{-(N-131)^{2}/3}]\\
&+c_{6}e^{-[(Z-51.5)^{2}+(N-80.5)^{2}]/1.9}
\end{array}
\end{equation}
is the shell correction term. Z and N are the proton number and
neutron number of the parent nuclei. $T_{1/2}$ is the half-life of
$\beta^{-}$-decay (in seconds). The parameters are $c_{1} =
3.37\times10^{-4}$, $c_{2} = -0.2558$, $c_{3} = 0.4028$, $c_{4} =
-1.0100$, $c_{5} = 0.9039$, and $c_{6} = 7.7139$.

For $\beta^{+}$-decay, in Ref.~\cite{Zha2}, Zhang \emph{et al}.
proposed a similar formula to the Eq. (10). It is written as:
\begin{equation}
\texttt{log}_{10}T_{1/2}=(c_{1}Z+c_{2})N+c_{3}Z+c_{4}.
\end{equation}
For different order (the allowed $\beta^{+}$-transition, the first
and the second forbidden $\beta^{+}$-transition), the parameters are
different. The even-odd effect has been taken into account in the
above equation. The best fit parameters are displayed in table 2.
\end{multicols}
\begin{center}
\tabcaption{ \label{tab1} The parameters of the Eq. (12). The word \lq\lq
order\rq\rq\;in the first column denotes the order of the
$\beta^{+}$-decay from ground state to ground state. The even-odd
effect has been included.}
\footnotesize
\begin{tabular*}{100mm}{c c c c c c c }
\hline
order&$c_{1}$&$c_{2}$&$c_{3}$&\multicolumn {3}{c}{$c_{4}$}\\
&&&&e-o, o-e&o-o&e-e\\
\hline
allowed&$-$0.00179&0.4233&$-$0.3405&$-$0.6443&$-$1.7089&$-$0.2132\\
  first&$-$0.00127&0.3992&$-$0.4183&3.8215   &3.7969   &4.0364\\
 second&$-$0.00162&0.3980&$-$0.3286&$-$0.1618&$-$0.4854&0.0267\\
\hline
\end{tabular*}
\end{center}
\begin{multicols}{2}
For a given proton number Z, we select ten continuous isotopes
nearest to the top and bottom of the $\beta$-stable region,
respectively. Thus there are 20 nuclei for each isotopic chain.
Because only the half-lives of the allowed $\beta^{+}$-transition,
the first and the second forbidden $\beta^{+}$-transition can be
calculated by the Eq. (12), the nuclei with higher forbidden
$\beta^{+}$-transition are not included. Because the formula (9) can
only calculate the half-lives of the nuclei with Z $\geq$ 84 and N
$\geq$ 128, the nuclei with N $<$ 128 are not included also. So the
number of the calculated nuclei of each isotopic chain may be less
than 20. We calculate the half-lives of the nuclei from Z = 90 to Z
= 126, and predict the decay modes of them. Because the calculated
data are too many, we firstly compare the calculated results with
the available experimental data~\cite{Audi2}. The selected
region for comparison is from Z = 90 to Z = 103, because there are
many experimental data in this region. The results are listed in
table 3.

\end{multicols}

\begin{footnotesize}
\begin{longtable}{ccrrccrr}
\caption[!htp]{ The comparison of the half-lives and decay modes
between the calculated results and the experimental data by Audi
\emph{et al}.~\cite{Audi2} from Z = 90 to Z = 103. Here
$C=\texttt{log}_{10}(T^{cal}_{1/2}/T^{expt}_{1/2})$. }
\vspace{-0.2cm}\\
\hline \hline  $\begin{array}{c} $$\\Z\\$$
\end{array}$ & $\begin{array}{c} $$\\A\\$$
\end{array}$ & $\begin{array}{c} $Cal.$\\T^{cal}_{\alpha\,1/2}\\$(s)$
\end{array}$ & $\begin{array}{c} $Cal.$\\T^{cal}_{\beta\,1/2}\\$(s)$
\end{array}$ & $\begin{array}{c} $Calculated$\\$decay modes$\\$and intensities(\%)$
\end{array}$ & $\begin{array}{c} $Experimental$\\$decay modes$\\$and intensities(\%)$
\end{array}$ & $\begin{array}{c} $$\\T^{expt}_{1/2}\\$(s)$
\end{array}$ & $\begin{array}{c} $$\\C\\$$
\end{array}$ \\
\hline
\endfirsthead
\multicolumn{6}{c}%
{{\bfseries \tablename\ \thetable{} -- continued from previous page}} \\
\hline \hline $\begin{array}{c} $$\\Z\\$$
\end{array}$ & $\begin{array}{c} $$\\A\\$$
\end{array}$ & $\begin{array}{c} $Cal.$\\T^{cal}_{\alpha\,1/2}\\$(s)$
\end{array}$ & $\begin{array}{c} $Cal.$\\T^{cal}_{\beta\,1/2}\\$(s)$
\end{array}$ & $\begin{array}{c} $Calculated$\\$decay modes$\\$and intensities(\%)$
\end{array}$ & $\begin{array}{c} $Experimental$\\$decay modes$\\$and intensities(\%)$
\end{array}$ & $\begin{array}{c} $$\\T^{expt}_{1/2}\\$(s)$
\end{array}$ & $\begin{array}{c} $$\\C\\$$
\end{array}$ \\
\hline
\endhead
\hline \multicolumn{6}{r}{{(Continued on next page)}} \\
\endfoot
\hline
\endlastfoot
\hline
90&218&9.14$\times10^{-8}$&718.8 &$\alpha=100$&$\alpha=100$&1.09$\times10^{-7}$&$-$0.08\\
90&219&3.11$\times10^{-6}$&844.5 &$\alpha=100$&$\alpha=100$&1.05$\times10^{-6}$&0.47\\
90&221&5.67$\times10^{-4}$&3136.2&$\alpha=100$&$\alpha=100$&1.68$\times10^{-3}$&$-$0.47\\
90&222&0.0028             &9912.9&$\alpha=100$&$\alpha=100$&2.05$\times10^{-3}$&0.14\\
90&223&0.93  &1.16$\times10^{4}$ &$\alpha=100$&$\alpha=100$&0.6&0.19\\
90&231&1.63$\times10^{17}$&2823.1&$\beta^{-}=100$&$\beta^{-}=100$&9.19$\times10^{4}$&$-$1.51\\
90&233&2.13$\times10^{21}$&999.5 &$\beta^{-}=100$&$\beta^{-}=100$&1338&$-$0.13\\
90&234&2.02$\times10^{21}$&594.7 &$\beta^{-}=100$&$\beta^{-}=100$&2.08$\times10^{6}$&$-$3.54\\
90&235&3.78$\times10^{24}$&353.9 &$\beta^{-}=100$&$\beta^{-}=100$&432&$-$0.09\\
90&236&2.44$\times10^{24}$&210.6 &$\beta^{-}=100$&$\beta^{-}=100$&2250&$-$1.03\\
90&237&4.61$\times10^{29}$&125.3 &$\beta^{-}=100$&$\beta^{-}=100$&288&$-$0.36\\
90&238&2.87$\times10^{29}$&74.6  &$\beta^{-}=100$&$\beta^{-}=100$&564&$-$0.88\\
91&219&2.27$\times10^{-7}$&115.0 &$\alpha=100$&$\alpha=100$&5.30$\times10^{-8}$&0.63\\
91&220&2.94$\times10^{-6}$&208.9 &$\alpha=100$&$\alpha=100$&7.80$\times10^{-7}$&0.58\\
91&221&2.02$\times10^{-5}$&424.7 &$\alpha=100$&$\alpha=100$&5.90$\times10^{-6}$&0.53\\
91&223&0.0063             &1568.1&$\alpha=100$&$\alpha=100$, $\beta^{+}<0.001\#$&5.10$\times10^{-3}$&0.09\\
91&225&6.5                &5789.2&$\alpha=100$&$\alpha=100$&1.7&0.58\\
91&226&694.0 &1.05$\times10^{4}$&$\alpha=94$, $\beta^{+}=6$&$\alpha=74$, $\beta^{+}=26$&108&0.78\\
91&233&7.47$\times10^{16}$&4741.5&$\beta^{-}=100$&$\beta^{-}=100$&2.33$\times10^{6}$&$-$2.69\\
91&234&1.15$\times10^{19}$&2823.4&$\beta^{-}=100$&$\beta^{-}=100$&2.41$\times10^{4}$&$-$0.93\\
91&235&3.53$\times10^{19}$&1681.3&$\beta^{-}=100$&$\beta^{-}=100$&1470&0.06\\
91&236&1.35$\times10^{22}$&1001.2&$\beta^{-}=100$&$\beta^{-}=100$&546&0.26\\
91&237&1.17$\times10^{22}$&596.2 &$\beta^{-}=100$&$\beta^{-}=100$&522&0.06\\
91&238&7.10$\times10^{27}$&355.0 &$\beta^{-}=100$&$\beta^{-}=100$&136&0.42\\
91&239&8.19$\times10^{27}$&211.4 &$\beta^{-}=100$&$\beta^{-}=100$&6.48$\times10^{3}$&$-$1.49\\
92&223&3.83$\times10^{-4}$&212.4 &$\alpha=100$&$\alpha\approx100$, $\beta^{+}=0.2\#$&2.10$\times10^{-5}$&1.26\\
92&225&0.18               &779.4 &$\alpha=100$&$\alpha=100$&0.061&0.47\\
92&226&0.34               &2449.3&$\alpha=100$&$\alpha=100$&0.269&0.10\\
92&227&99.1&2860.9&$\alpha=97$, $\beta^{+}=3$&$\alpha=100$, $\beta^{+}<0.001\#$&66&0.16\\
92&228&659.4             &4508.8 &$\alpha=87$, $\beta^{+}=13$&$\alpha>95$, $\varepsilon<5$&546&0.02\\
92&229&8.59$\times10^{4}$ &2888.3&$\alpha=3$, $\beta^{+}=97$&$\alpha\approx20$, $\beta^{+}\approx80$&3.48$\times10^{3}$&$-$0.10\\
\emph{\textbf{92}}&\emph{\textbf{236}}&3.37$\times10^{15}$&4753.1&$\beta^{-}=100$&$\alpha=100$&7.39$\times10^{14}$&\\
92&237&1.62$\times10^{18}$&2832.6&$\beta^{-}=100$&$\beta^{-}=100$&5.83$\times10^{5}$&$-$2.30\\
92&239&1.62$\times10^{21}$&1006.0&$\beta^{-}=100$&$\beta^{-}=100$&1.41$\times10^{3}$&$-$0.15\\
92&240&8.57$\times10^{20}$&599.5 &$\beta^{-}=100$&$\beta^{-}=100$&5.08$\times10^{4}$&$-$1.92\\
92&242&4.35$\times10^{23}$&212.9 &$\beta^{-}=100$&$\beta^{-}=100$&1.01$\times10^{3}$&$-$0.68\\
93&227&1.34               &385.2 &$\alpha=100$&$\alpha\approx100$, $\beta^{+}=0.05\#$&0.51&0.42\\
93&228&118.8              &695.3 &$\alpha=85$, $\beta^{+}=15$&$\alpha=40$, $\varepsilon=60$&61.4&0.22\\
93&229&1033.1             &1405.5&$\alpha=58$, $\beta^{+}=42$&$\alpha>50$, $\beta^{+}<50$&240&0.39\\
93&230&3.00$\times10^{4}$ &64.6  &$\beta^{+}=100$&$\beta^{+}\leq97$, $\alpha\geq3$&276&$-$0.63\\
93&231&4.85$\times10^{5}$ &1354.4&$\beta^{+}=100$&$\beta^{+}=98$, $\alpha=2$&2.93$\times10^{3}$&$-$0.34\\
93&239&1.23$\times10^{16}$&4779.5&$\beta^{-}=100$&$\beta^{-}=100$&2.04$\times10^{5}$&$-$1.63\\
93&240&6.17$\times10^{16}$&2850.5&$\beta^{-}=100$&$\beta^{-}=100$&3.71$\times10^{3}$&$-$0.11\\
93&241&8.45$\times10^{18}$&1700.1&$\beta^{-}=100$&$\beta^{-}=100$&834&0.31\\
93&242&2.05$\times10^{19}$&1013.9&$\beta^{-}=100$&$\beta^{-}=100$&132&0.86\\
93&243&4.96$\times10^{19}$&604.7 &$\beta^{-}=100$&$\beta^{-}=100$&111&0.74\\
93&244&3.11$\times10^{21}$&360.7 &$\beta^{-}=100$&$\beta^{-}=100$&137&0.42\\
\emph{\textbf{94}}&\emph{\textbf{229}}&24.3&189.2 &$\alpha=89$, $\beta^{+}=11$&$\alpha=100$&120&\\
94&231&1.78$\times10^{4}$ &194.6 &$\alpha=1$, $\beta^{+}=99$&$\alpha=13$, $\beta^{+}=87$&516&$-$0.43\\
94&233&1.10$\times10^{6}$ &629.9 &$\beta^{+}=100$&$\beta^{+}\approx100$, $\alpha=0.12$&1.25$\times10^{3}$&$-$0.30\\
94&235&1.86$\times10^{8}$ &2038.6&$\beta^{+}=100$&$\beta^{+}\approx100$, $\alpha=0.003$&1.52$\times10^{3}$&0.13\\
\emph{\textbf{94}}&\emph{\textbf{242}}&1.17$\times10^{13}$&4821.1&$\beta^{-}=100$&$\alpha=100$&4.93$\times10^{11}$&\\
94&243&2.30$\times10^{15}$&2877.5&$\beta^{-}=100$&$\beta^{-}=100$&1.78$\times10^{4}$&$-$0.79\\
\emph{\textbf{94}}&\emph{\textbf{244}}&1.85$\times10^{15}$&1717.5&$\beta^{-}=100$&$\alpha\approx100$, SF =0.12&1.05$\times10^{14}$&\\
94&245&2.50$\times10^{17}$&1025.1&$\beta^{-}=100$&$\beta^{-}=100$&3.78$\times10^{4}$&$-$1.57\\
94&246&4.72$\times10^{17}$&611.9 &$\beta^{-}=100$&$\beta^{-}=100$&9.36$\times10^{5}$&$-$3.18\\
94&247&1.00$\times10^{20}$&365.2 &$\beta^{-}=100$&$\beta^{-}=100$&1.96$\times10^{5}$&$-$2.73\\
95&235&3.10$\times10^{5}$ &1202.3&$\beta^{+}=100$&$\beta^{+}\approx100$, $\alpha=0.4$&594&0.31\\
95&237&2.29$\times10^{7}$ &932.6 &$\beta^{+}=100$&$\beta^{+}\approx100$, $\alpha=0.025$&4.38$\times10^{3}$&$-$0.67\\
95&245&5.58$\times10^{12}$&4878.1&$\beta^{-}=100$&$\beta^{-}=100$&7.38$\times10^{3}$&$-$0.18\\
95&246&4.73$\times10^{13}$&2913.8&$\beta^{-}=100$&$\beta^{-}=100$&2.34$\times10^{3}$&0.10\\
95&247&1.27$\times10^{15}$&1740.5&$\beta^{-}=100$&$\beta^{-}=100$&1.38$\times10^{3}$&0.10\\
96&239&1.58$\times10^{6}$ &927.3 &$\beta^{+}=100$&$\beta^{+}\approx100$, $\alpha<0.1$&1.04$\times10^{4}$&$-$1.05\\
96&249&2.21$\times10^{13}$&2959.7&$\beta^{-}=100$&$\beta^{-}=100$&3.85$\times10^{3}$&$-$0.11\\
96&251&1.89$\times10^{13}$&1057.7&$\beta^{-}=100$&$\beta^{-}=100$&1.01$\times10^{3}$&0.02\\
97&242&5.02$\times10^{5}$ &1750.9&$\beta^{+}=100$&$\beta^{+}\approx100$&420&0.62\\
97&243&1.65$\times10^{5}$ &3498.3&$\alpha=2$, $\beta^{+}=98$&$\alpha\approx0.15$, $\beta^{+}\approx100$&1.62$\times10^{4}$&$-$0.67\\
97&251&1.38$\times10^{11}$&5040.9&$\beta^{-}=100$&$\beta^{-}=100$&3.34$\times10^{3}$&0.18\\
98&241&522.5              &130.6 &$\alpha=20$, $\beta^{+}=80$&$\alpha\approx25$, $\beta^{+}\approx75$&228&$-$0.34\\
98&243&8779.7             &462.8 &$\alpha=5$, $\beta^{+}=95$&$\alpha\approx16$, $\beta^{+}\approx84$&642&$-$0.17\\
98&245&1.66$\times10^{4}$ &1640.1&$\alpha=9$, $\beta^{+}=91$&$\alpha=36$, $\beta^{+}=64$&2.70$\times10^{3}$&$-$0.26\\
\emph{\textbf{98}}&\emph{\textbf{246}}&1.22$\times10^{5}$ &5064.3&$\alpha=4$, $\beta^{+}=96$&$\alpha=100$&1.28$\times10^{5}$&\\
98&253&1.69$\times10^{9}$ &8598.6&$\beta^{-}=100$&$\beta^{-}\approx100$, $\alpha\approx0.31$&1.54$\times10^{6}$&$-$2.25\\
98&255&1.84$\times10^{11}$&3082.3&$\beta^{-}=100$&$\beta^{-}=100$&5.10$\times10^{3}$&$-$0.22\\
99&243&33.4               &61.6  &$\alpha=65$, $\beta^{+}=35$&$\alpha\geq30$, $\beta^{+}\leq70$&21&0.01\\
99&245&120.3              &217.0 &$\alpha=64$, $\beta^{+}=36$&$\alpha=40$, $\beta^{+}=60$&66&0.07\\
99&247&3892.3             &116.8 &$\alpha=3$, $\beta^{+}=97$&$\alpha\approx7$, $\beta^{+}\approx93$&276&$-$0.39\\
99&248&2.65$\times10^{5}$ &1355.9&$\beta^{+}=100$&$\beta^{+}\approx100$, $\alpha\approx0.25$&1620&$-$0.08\\
99&249&5.97$\times10^{5}$ &2693.4&$\beta^{+}=100$&$\beta^{+}\approx100$, $\alpha\approx0.57$&6.13$\times10^{3}$&$-$0.36\\
99&256&3.72$\times10^{9}$ &8802.2&$\beta^{-}=100$&$\beta^{-}=100$&1.52$\times10^{3}$&0.76\\
99&257&8.50$\times10^{9}$ &5274.1&$\beta^{-}=100$&$\beta^{-}=100$&6.65$\times10^{5}$&$-$2.10\\
100&247&37.2              &16.5  &$\alpha=31$, $\beta^{+}=69$&$\alpha\geq50$, $\beta^{+}\leq50$&35&$-$0.49\\
100&248&34.3              &310.5 &$\alpha=90$, $\beta^{+}=10$&$\alpha=93$, $\beta^{+}=7$&36&$-$0.07\\
100&251&2.39$\times10^{4}$&411.3 &$\alpha=2$, $\beta^{+}=98$&$\alpha=1.8$, $\beta^{+}=98.2$&1.91$\times10^{4}$&$-$1.67\\
101&250&106.6             &82.6  &$\alpha=44$, $\beta^{+}=56$&$\alpha=7$, $\beta^{+}=93$&52&$-$0.05\\
\emph{\textbf{101}}&\emph{\textbf{252}}&2655.3 &3.30  &$\beta^{+}=100$&$\beta^{+}>50$, $\alpha<50$&138&\\
102&254&33.3              &228.1 &$\alpha=87$, $\beta^{+}=13$&$\alpha=90$, $\beta^{+}=10$&51&$-$0.25\\
102&256&2.33              &789.7 &$\alpha=100$&$\alpha\approx100$&2.91&$-$0.10\\
103&253&1.68              &1.30  &$\alpha=44$, $\beta^{+}=56$&$\alpha=90$, $\beta^{+}=1\#$&0.58&$-$0.01\\
103&254&16.8              &17.3  &$\alpha=51$, $\beta^{+}=49$&$\alpha=76$, $\beta^{+}=24$&13&$-$0.19\\
103&256&13.6              &0.60  &$\alpha=4$, $\beta^{+}=96$&$\alpha=85$, $\beta^{+}=15$&27&$-$1.68\\
\hline
\end{longtable}
\end{footnotesize}

\begin{multicols}{2}

The calculated half-lives are obtained from the formulae (9), (10)
and (12). When using the formula (9) to calculate half-lives of
$\alpha$-decay and $\beta^{+}$-decay, we need $\alpha$-decay
energies $Q_{\alpha}$, spins and parities. Here, All the values of
$Q_{\alpha}$ are taken from Ref.~\cite{Audi2, Audi}. If there are no
experimental data, we use the calculated data obtained by M\"{o}ller
\emph{et al}. ~\cite{Moll}. The calculated $\alpha$-decay
half-lives and the calculated $\beta$-decay (including
$\beta^{-}$-decay and $\beta^{+}$-decay) half-lives (in seconds) are
listed in the third and the fourth columns in table 3. The fifth
column is the calculated decay modes and intensities (in \%). The
decay mode can be regard as a competition between $\alpha$-decay and
$\beta$-decay. Here we define a symbol $R$ to denote the ratio of
$\alpha$-decay half-life and $\beta$-decay half-life. It is defined
as: $R=T^{cal}_{\alpha\; 1/2}/T^{cal}_{\beta \;1/2}$. If the
$\alpha$-decay half-life is shorter than the $\beta$-decay half-life
by 100 times (i.e. $R<0.01$) in a nucleus, we can say the decay mode
of this nucleus is $\alpha$-decay. If the $\alpha$-decay half-life
is longer than the $\beta$-decay half-life by 100 times (i.e.
$R>100$) in a nucleus, we can say the decay mode of this nucleus is
$\beta$-decay. If $0.01<R<100$, the decay mode can be regard as a
coexistence state of both $\alpha$-decay and $\beta$-decay, and we
use the symbol $\alpha +\beta^{+}$ to denote the coexistence state
of both $\alpha$-decay and $\beta^{+}$-decay. The sixth and seventh
columns are the experimental decay modes and intensities (in \%) and
half-lives. The data marked with \# denote the values from
systematic trends in neighboring nuclei. The symbol $C$ in the last
column is the ratio of calculated half-life and experimental one,
and it is in the form of
$C=\texttt{log}_{10}(T^{cal}_{1/2}/T^{expt}_{1/2})$. The nuclei
which do not have explicit experimental decay modes and intensities
(in \%) are not included in table 3. There are altogether 91 nuclei.
It can be seen that the predicted decay modes are in excellent
agreement with the experimental ones. There are only six nuclei
whose decay modes are not in line with the predicted decay modes. The proton numbers and the mass numbers of these six
nuclei are marked in bold italic type in table 3. It indicates that the
predicted decay modes in this region are reliable. For the ratios of
calculated half-lives and experimental ones, the values of $C$ are
mostly between 1.0 and $-1.0$. This means that the most differences
between calculated half-lives and experimental ones are less than 10
times. It is a good approximation. Usually, if the differences
between theoretical half-lives and experimental ones are less than
$10^{3}$ times, it can be said that the results are satisfactory. So
we can say the calculated half-lives in this region are in good
agreement with the experimental data.

After having compared the calculated results with the experimental
data from Z = 90 to Z = 103, we predict some half-lives and decay
modes for some heavier proton number Z. Synthesizing superheavy
nuclei is a hot topic of nuclear physics. At present, Chinese
physicists are endeavoring to synthesize some superheavy nuclei
around Z = 110, and they have obtained some successful results
~\cite{Gan}. Next, we select the region from Z = 107 to Z = 110
to make some predictions. There are many researches in this region
~\cite{Gan, Oga1, Oga3, Dull, Gupt, Drag, Pei, Dvor, Zuo, Mori}.
We hope that our predictions will be useful for future experiments
on heavy and superheavy nuclei. Because there are almost no explicit
spins and parities for the nuclei with Z $\geq$ 107 except even-even
nuclei, it is difficult for us to judge the orders of
$\beta^{+}$-decay. When using the formula (12) to calculate
half-lives of $\beta^{+}$-decay, we suppose all the orders of
$\beta^{+}$-decay are the first $\beta^{+}$-transition for
simplifying the calculations. The calculated results are listed in
table 4.
\end{multicols}
\begin{footnotesize}
\begin{longtable}[!htp]{cccccrc}
\caption[htp]{ The calculated half-lives and the predicted decay
modes of the nuclei from Z = 107 to Z = 110. Some available experimental
half-lives of $\alpha$-decay ~\cite{Gan, Oga1, Gupt, Mori} are
listed for comparison. Here $D=T^{cal}_{\alpha\,1/2}/T^{expt}_{\alpha\,1/2}$.}
\label{tab5}\vspace{-0.2cm}\\
\hline \hline $\begin{array}{c} $$\\Z\\$$
\end{array}$ & $\begin{array}{c} $$\\A\\$$
\end{array}$ & $\begin{array}{c} $Cal.$\\T^{cal}_{\alpha\,1/2}\\$(s)$
\end{array}$ & $\begin{array}{c} $Cal.$\\T^{cal}_{\beta\,1/2}\\$(s)$
\end{array}$ & $\begin{array}{c} $Calculated$\\$decay modes$\\$and intensities(\%)$
\end{array}$ & $\begin{array}{c} $$\\T^{expt}_{\alpha\,1/2}\\$(s)$
\end{array}$ & $\begin{array}{c} $$\\D\\$$
\end{array}$ \\
\hline
\endfirsthead
\multicolumn{6}{c}%
{{\bfseries \tablename\ \thetable{} -- continued from previous page}} \\
\hline \hline $\begin{array}{c} $$\\Z\\$$
\end{array}$ & $\begin{array}{c} $$\\A\\$$
\end{array}$ & $\begin{array}{c} $Cal.$\\T^{cal}_{\alpha\,1/2}\\$(s)$
\end{array}$ & $\begin{array}{c} $Cal.$\\T^{cal}_{\beta\,1/2}\\$(s)$
\end{array}$ & $\begin{array}{c} $Calculated$\\$decay modes$\\$and intensities(\%)$
\end{array}$ & $\begin{array}{c} $$\\T^{expt}_{\alpha\,1/2}\\$(s)$
\end{array}$ & $\begin{array}{c} $$\\D\\$$
\end{array}$ \\
\hline
\endhead
\hline \multicolumn{6}{r}{{(Continued on next page)}} \\
\endfoot
\hline
\endlastfoot
\hline
107&264&0.16               &2.40              &$\alpha=94$, $\beta^{+}=6$&0.9 \cite{Mori}&0.18\\
107&265&1.99               &4.64              &$\alpha=70$, $\beta^{+}=30$&0.94 \cite{Gan}&2.12\\
107&266&20.9               &8.04              &$\alpha=28$, $\beta^{+}=72$&5 \cite{Audi2}&4.18\\
107&267&36.3               &15.6              &$\alpha=30$, $\beta^{+}=70$&17 \cite{Gupt}&2.14\\
107&268&55.3               &27                &$\alpha=33$, $\beta^{+}=67$&&\\
107&269&86.5               &52.4              &$\alpha=38$, $\beta^{+}=62$&&\\
107&270&63.7               &90.9              &$\alpha=59$, $\beta^{+}=41$&61 \cite{Oga1}&1.04\\
107&271&0.91               &176.3             &$\alpha=100$&&\\
107&272&34.1               &305.5             &$\alpha=90$, $\beta^{+}=10$&9.8 \cite{Oga1}&3.48\\
107&273&56.2               &$\beta$-stable    &$\alpha=100$&&\\
107&274&3.99$\times10^{3}$ &$\beta$-stable    &$\alpha=100$&&\\
107&275&108.5              &$\beta$-stable    &$\alpha=100$&&\\
107&276&4.00$\times10^{4}$ &$\beta$-stable    &$\alpha=100$&&\\
107&277&4.58$\times10^{6}$ &$\beta$-stable    &$\alpha=100$&&\\
107&278&9.12$\times10^{5}$ &$\beta$-stable    &$\alpha=100$&&\\
107&279&1.07$\times10^{7}$ &$\beta$-stable    &$\alpha=100$&&\\
107&280&2.49$\times10^{10}$&1.19$\times10^{4}$&$\beta^{-}=100$&&\\
107&281&3.24$\times10^{11}$&7.16$\times10^{3}$&$\beta^{-}=100$&&\\
107&282&3.42$\times10^{13}$&4.31$\times10^{3}$&$\beta^{-}=100$&&\\
107&283&1.52$\times10^{12}$&2.60$\times10^{3}$&$\beta^{-}=100$&&\\
107&284&1.16$\times10^{12}$&1.57$\times10^{3}$&$\beta^{-}=100$&&\\
107&285&9.45$\times10^{11}$&945.6             &$\beta^{-}=100$&&\\
107&286&4.41$\times10^{15}$&570.1             &$\beta^{-}=100$&&\\
107&287&5.48$\times10^{14}$&343.7             &$\beta^{-}=100$&&\\
107&288&7.69$\times10^{15}$&207.2             &$\beta^{-}=100$&&\\
107&289&1.43$\times10^{15}$&124.9             &$\beta^{-}=100$&&\\
108&266&2.72$\times10^{-3}$&1.83              &$\alpha=100$&2.3$\times10^{-3}$ \cite{Gupt}&1.18\\
108&267&0.129              &2.04              &$\alpha=94$, $\beta^{+}=6$&0.058 \cite{Gupt}&2.22\\
108&268&0.0376             &6.12              &$\alpha=100$&&\\
108&269&9.44               &6.82              &$\alpha=42$, $\beta^{+}=58$&9.7 \cite{Gupt}&0.97\\
108&270&1.88               &20.4              &$\alpha=92$, $\beta^{+}=8$&3.6 \cite{Gupt}&0.52\\
108&271&0.21               &22.8              &$\alpha=100$&&\\
108&272&0.011              &68.3              &$\alpha=100$&&\\
108&273&0.21               &76.2              &$\alpha=100$&&\\
108&274&0.49               &228.4             &$\alpha=100$&&\\
108&275&0.31               &254.6             &$\alpha=98$, $\beta^{+}=2$&0.19 \cite{Oga1}&1.63\\
108&276&66.3               &$\beta$-stable    &$\alpha=100$&&\\
108&277&8.00$\times10^{3}$ &$\beta$-stable    &$\alpha=100$&&\\
108&278&516.2              &$\beta$-stable    &$\alpha=100$&&\\
108&279&9.31$\times10^{4}$ &$\beta$-stable    &$\alpha=100$&&\\
108&280&1.98$\times10^{6}$ &$\beta$-stable    &$\alpha=100$&&\\
108&281&4.57$\times10^{10}$&$\beta$-stable    &$\alpha=100$&&\\
108&282&1.60$\times10^{10}$&$\beta$-stable    &$\alpha=100$&&\\
108&283&3.50$\times10^{11}$&1.25$\times10^{4}$&$\beta^{-}=100$&&\\
108&284&1.60$\times10^{10}$&7.54$\times10^{3}$&$\beta^{-}=100$&&\\
108&285&1.71$\times10^{10}$&4.55$\times10^{3}$&$\beta^{-}=100$&&\\
108&286&8.25$\times10^{9}$ &2.48$\times10^{3}$&$\beta^{-}=100$&&\\
108&287&9.52$\times10^{13}$&1.66$\times10^{3}$&$\beta^{-}=100$&&\\
108&288&4.23$\times10^{12}$&999.3             &$\beta^{-}=100$&&\\
108&289&1.09$\times10^{14}$&602.9             &$\beta^{-}=100$&&\\
108&290&2.22$\times10^{13}$&363.8             &$\beta^{-}=100$&&\\
108&291&1.09$\times10^{14}$&219.5             &$\beta^{-}=100$&&\\
108&292&4.83$\times10^{12}$&132.5             &$\beta^{-}=100$&&\\
109&269&7.33$\times10^{-3}$&0.891             &$\alpha=100$&&\\
109&270&3.62$\times10^{-4}$&1.54              &$\alpha=88$, $\beta^{+}=12$&5$\times10^{-3}$ \cite{Gupt}&0.07\\
109&271&0.073              &2.96              &$\alpha=98$, $\beta^{+}=2$&&\\
109&272&0.018              &5.1               &$\alpha=100$&&\\
109&273&1.45$\times10^{-3}$&9.84              &$\alpha=100$&&\\
109&274&1.14               &17                &$\alpha=94$, $\beta^{+}=6$&0.445 \cite{Oga1}&2.56\\
109&275&9.82$\times10^{-3}$&32.7              &$\alpha=100$&9.7$\times10^{-3}$ \cite{Oga1}&1.01\\
109&276&1.57               &56.3              &$\alpha=97$, $\beta^{+}=3$&0.72 \cite{Oga1}&2.18\\
109&277&4.28               &108.9             &$\alpha=96$, $\beta^{+}=4$&&\\
109&278&240.6              &187.2             &$\alpha=44$, $\beta^{+}=56$&&\\
109&279&1.29$\times10^{3}$ &$\beta$-stable    &$\alpha=100$&&\\
109&280&3.21$\times10^{4}$ &$\beta$-stable    &$\alpha=100$&&\\
109&281&1.90$\times10^{5}$ &$\beta$-stable    &$\alpha=100$&&\\
109&282&1.94$\times10^{9}$ &$\beta$-stable    &$\alpha=100$&&\\
109&283&2.06$\times10^{9}$ &$\beta$-stable    &$\alpha=100$&&\\
109&284&1.70$\times10^{10}$&$\beta$-stable    &$\alpha=100$&&\\
109&285&8.22$\times10^{8}$ &$\beta$-stable    &$\alpha=100$&&\\
109&286&4.41$\times10^{8}$ &$\beta$-stable    &$\alpha=100$&&\\
109&287&1.69$\times10^{9}$ &7.97$\times10^{3}$&$\beta^{-}=100$&&\\
109&288&1.40$\times10^{12}$&4.81$\times10^{3}$&$\beta^{-}=100$&&\\
109&289&2.02$\times10^{12}$&2.90$\times10^{3}$&$\beta^{-}=100$&&\\
109&290&1.47$\times10^{13}$&1.75$\times10^{3}$&$\beta^{-}=100$&&\\
109&291&7.61$\times10^{12}$&1.06$\times10^{3}$&$\beta^{-}=100$&&\\
109&292&1.88$\times10^{13}$&639.6             &$\beta^{-}=100$&&\\
109&293&1.00$\times10^{12}$&386.3             &$\beta^{-}=100$&&\\
109&294&1.62$\times10^{9}$ &233.2             &$\beta^{-}=100$&&\\
109&295&9.23$\times10^{8}$ &140.8             &$\beta^{-}=100$&&\\
109&296&1.40$\times10^{10}$&85.1              &$\beta^{-}=100$&&\\
110&272&1.02$\times10^{-4}$&1.15              &$\alpha=100$&&\\
110&273&2.16$\times10^{-4}$&1.28              &$\alpha=100$&1.7$\times10^{-4}$ \cite{Gupt}&1.27\\
110&274&3.32$\times10^{-5}$&3.81              &$\alpha=100$&&\\
110&275&8.90$\times10^{-4}$&4.23                 &$\alpha=100$&&\\
110&276&2.53$\times10^{-3}$&12.6              &$\alpha=100$&&\\
110&277&0.081              &14.0              &$\alpha=100$&&\\
110&278&0.091              &41.6              &$\alpha=100$&&\\
110&279&1.40               &46.1              &$\alpha=97$, $\beta^{+}=3$&2 \cite{Oga1}&0.7\\
110&280&9.18               &137.5             &$\alpha=94$, $\beta^{+}=6$&&\\
110&281&584.3              &152.4             &$\alpha=21$, $\beta^{+}=79$&240 \cite{Audi2}&2.43\\
110&282&1.94$\times10^{9}$ &$\beta$-stable    &$\alpha=100$&&\\
110&283&3.97$\times10^{9}$ &$\beta$-stable    &$\alpha=100$&&\\
110&284&1.14$\times10^{9}$ &$\beta$-stable    &$\alpha=100$&&\\
110&285&1.13$\times10^{9}$ &$\beta$-stable    &$\alpha=100$&&\\
110&286&8.62$\times10^{9}$ &$\beta$-stable    &$\alpha=100$&&\\
110&287&5.29$\times10^{9}$ &$\beta$-stable    &$\alpha=100$&&\\
110&288&8.76$\times10^{9}$ &$\beta$-stable    &$\alpha=100$&&\\
110&289&3.36$\times10^{9}$ &$\beta$-stable    &$\alpha=100$&&\\
110&290&3.12$\times10^{10}$&8.45$\times10^{3}$&$\beta^{-}=100$&&\\
110&291&1.74$\times10^{11}$&5.10$\times10^{3}$&$\beta^{-}=100$&&\\
110&292&2.24$\times10^{10}$&3.08$\times10^{3}$&$\beta^{-}=100$&&\\
110&293&6.49$\times10^{10}$&1.86$\times10^{3}$&$\beta^{-}=100$&&\\
110&294&6.07$\times10^{9}$ &1.13$\times10^{3}$&$\beta^{-}=100$&&\\
110&295&7.67$\times10^{6}$ &680.7             &$\beta^{-}=100$&&\\
110&296&6.59$\times10^{6}$ &411.4             &$\beta^{-}=100$&&\\
110&297&7.84$\times10^{7}$ &248.6             &$\beta^{-}=100$&&\\
110&298&2.38$\times10^{6}$ &150.2             &$\beta^{-}=100$&&\\
110&299&2.26$\times10^{8}$ &90.8              &$\beta^{-}=100$&&\\
\end{longtable}
\end{footnotesize}
\begin{multicols}{2}

In table 4, the calculated half-lives and the suggested decay modes
of the nuclei from Z =107 to Z = 110 can be clearly seen. The
available experimental half-lives of $\alpha$-decay are listed in
the sixth column, and the corresponding references are also listed
there. The symbol $D$ in the last column is the ratio of calculated
half-life of $\alpha$-decay and experimental one, and it is in the
form of $D=T^{cal}_{\alpha \,1/2}/T^{expt}_{\alpha \,1/2}$.

For $\beta^{-}$-decay, the values of half-life vary from $10^{1}$ s
to $10^{5}$ s for all Z. The nearer the nuclei are close to the
$\beta$-stable region, the longer their half-lives are. For
$\beta^{+}$-decay, on the whole, it is similar to the case of
$\beta^{-}$-decay. For $\alpha$-decay, the half-lives of
$\alpha$-decay approximately vary from $10^{-4}$ s to $10^{16}$ s in
this region. The half-life of $\alpha$-decay of a nucleus above the
$\beta$-stable region is much longer than its half-life of
$\beta^{-}$-decay on the whole. Thus the nuclei above the
$\beta$-stable region occur $\beta^{-}$-decay mainly. However, for
most nuclei below the $\beta$-stable region, their half-lives of
$\alpha$-decay are slightly less than the ones of $\beta$-decay. So
the decay modes of these nuclei are mainly $\alpha$ or
$\alpha+\beta^{+}$-decay. There are 18 experimental half-lives of
$\alpha$-decay in this region. It can be seen that the calculated
half-lives of $\alpha$-decay are in agreement with the experimental
ones. Except for $^{270}$Mt ($D=0.07$), the values of $D$ vary from 0.18
to 4.18. It is a good approximation. It must be pointed out that the
half-life is very sensitive to the $\alpha$-decay energy. A small
change in $\alpha$-decay energy will lead to a very large difference
in half-life. There are few experimental $\alpha$-decay energies in
this region, and most $\alpha$-decay energies used for calculation
are the estimated data~\cite{Audi} or the calculated results~\cite{Moll}.

To clearly understand the competition between $\alpha$-decay and
$\beta$-decay of the nuclei close to the calculated $\beta$-stable
region, we draw the predicted decay modes from Z = 90 to Z = 126 in
figure 3.
\begin{center}
\includegraphics[width=8cm]{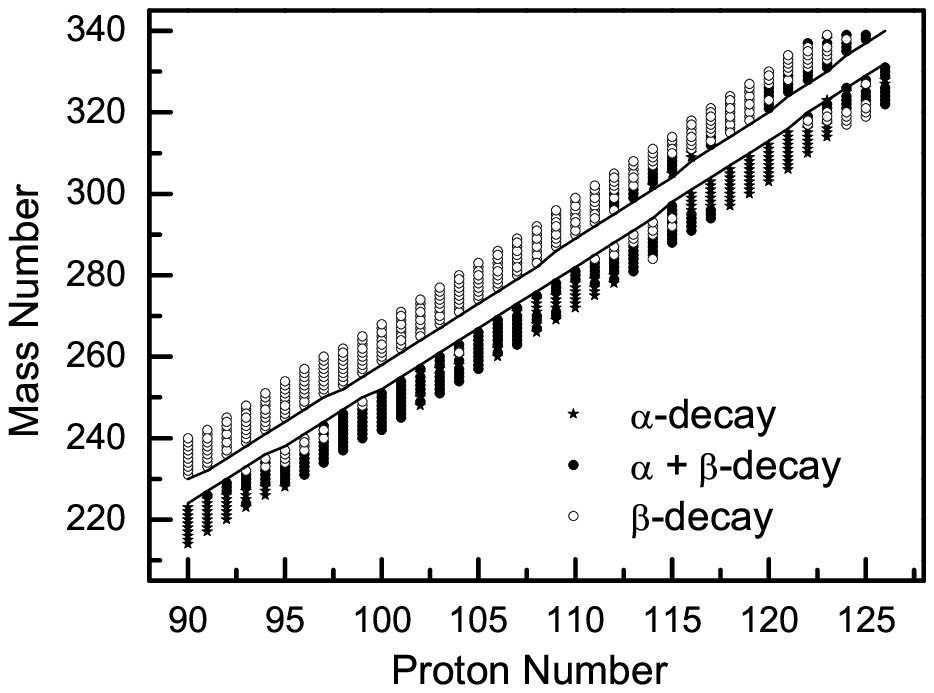}
\figcaption{\label{fig3} The
predicted decay modes of the nuclei close to the calculated
$\beta$-stable region. The dark asterisks denote $\alpha$-decay. The
dark circles denote $\alpha$ + $\beta$-decay. The hollow circles
denote $\beta$-decay. }
\end{center}

In figure 3, one can clearly see the decay modes of the nuclei close
to the calculated $\beta$-stable region. The dark asterisks denote
$\alpha$-decay. The dark circles denote $\alpha$ and $\beta$-decay.
The hollow circles denote $\beta$-decay. The decay modes are mostly
$\beta^{-}$-decay above the $\beta$-stable region. Especially for Z
$\leq$ 111, all the decay modes are $\beta^{-}$-decay. The decay
modes are very complex below the $\beta$-stable region. All the
three cases of decay mode can occur from Z = 90 to Z = 126. It
indicates that the competition between $\alpha$-decay and
$\beta$-decay is very complex and drastic below the $\beta$-stable
region. It can be seen that the nuclei above the $\beta$-stable
region can occur $\alpha$-decay and $\beta^{-}$-decay
($\alpha+\beta^{-}$) simultaneously when Z $\geq$ 112. It is a very
interesting phenomenon, because there is not the decay mode of
$\alpha+\beta^{-}$according to experimental results by Audi \emph{et
al}. for all Z.

\section{Conclusions}
In summary, we propose the $\beta$-stable region for Z $\geq$ 90. The predicted
$\beta$-stable nuclei in the calculated $\beta$-stable region are in
good agreement with the ones obtained by M\"{o}ller \emph{et al}.. We
calculate the half-lives of the nuclei close to the calculated
$\beta$-stable region and systematically study the competition between
$\alpha$-decay and $\beta$-decay. The calculated
half-lives and the suggested decay modes are in good agreement with
the experimental results from Audi's Table. The predictions for half-lives and decay modes of the
nuclei with Z = 107$-$110 are presented. We draw the predicted decay modes
from Z = 90 to Z = 126 in a figure. We find the nuclei above
the $\beta$-stable region can occur $\alpha$-decay and
$\beta^{-}$-decay ($\alpha+\beta^{-}$) simultaneously when Z $\geq$
112. It is a very interesting phenomenon. The competition between $\alpha$-decay and
$\beta$-decay is very complex and drastic below the $\beta$-stable
region. The calculated results on
the half-lives and the decay modes of the nuclei close to the
calculated $\beta$-stable region are useful for the future experiments on heavy and
superheavy nuclei.

\end{multicols}

\clearpage

\end{CJK*}
\end{document}